\documentclass[preprint,aps]{revtex4-1}
\usepackage{graphicx}
\usepackage{dcolumn}
\usepackage{bm}

\begin{document}

\title{Non-Markovian dynamics for an open two-level
system without rotating wave approximation: Indivisibility versus
backflow of information}

\author{Hao-Sheng Zeng\footnote{Corresponding author
email: hszeng@hunnu.edu.cn}, Ning Tang, Yan-Ping Zheng and
Tian-Tian Xu}
\address {Key Laboratory of Low-Dimensional Quantum
Structures and Quantum Control of Ministry of Education, and
Department of Physics, Hunan Normal University, Changsha
410081, China}
\date{\today}

\begin{abstract}

By use of the two measures presented recently, the indivisibility
and the backflow of information, we study the non-Markovianity of
the dynamics for a two-level system interacting with a
zero-temperature structured environment without using rotating wave
approximation (RWA). In the limit of weak coupling between the
system and the reservoir, and by expanding the time-convolutionless
(TCL) generator to the forth order with respect to the coupling
strength, the time-local non-Markovian master equation for the
reduced state of the system is derived. Under the secular
approximation, the exact analytic solution is obtained and the
sufficient and necessary conditions for the indivisibility and the
backflow of information for the system dynamics are presented. In
the more general case, we investigate numerically the properties of
the two measures for the case of Lorentzian reservoir. Our results
show the importance of the counter-rotating terms to the
short-time-scale non-Markovian behavior of the system dynamics,
further expose the relations between the two measures and their
rationality as non-Markovian measures. Finally, the complete
positivity of the dynamics of the considered system is discussed.
\end{abstract}

\maketitle

\section{Introduction}

Realistic quantum systems cannot avoid interactions with their
environments, thus the study of open quantum systems is very
important. It is not only relevant for better understanding of
quantum theory, but also fundamental for various modern applications
of quantum mechanics, especially for quantum communication,
cryptography and computation \cite{Nielsen}. The early study of
dynamics of open quantum systems usually consists in the application
of an appropriate Born-Markov approximation, that is, neglects all
the memory effects, leading to a master equation which can be cast
in the so-called Lindblad form \cite{Lindblad,Gorini}. Master
equations in Lindblad form can be characterized by the fact that the
dynamics of the system satisfies both the semigroup property and the
complete positivity, thus ensuring the preservation of positivity of
the density matrix during the time evolution. We usually attribute
the dynamical processes with these evolutional properties to the
well-known Markovian ones.

However, people recently found that Many relevant physical systems,
such as the quantum optical system \cite{Breuer3}, quantum dot
\cite{Kubota}, superconductor system \cite{Yinghua}, could not be
described simply by Markovian dynamics. Similarly, quantum chemistry
\cite{Shao} and the excitation transfer of a biological system
\cite{Chin} also need to be treated as non-Markovian processes.
Quantum non-Markovian processes can lead to distinctly different
effects on decoherence and disentanglement
\cite{Dijkstra,Anastopoulos} of open systems compared to Markovian
processes. These non-markovian effects can on the one hand 
enrich the basic theory of quantum mechanics, 
on the other hand benefit the quantum information processing. Because of these distinctive
properties and extensive applications, more and more attention and
interest have been devoted to the study of non-Markovian processes
of open systems, including the measures of non-Markovianity
\cite{Breuer,Laine,Rivas,Wolf,Usha,Lu,Hou,Xu,He}, the positivity
\cite{Breuer1,Shabani,Breuer2}, and some other dynamical properties
\cite{Haikka,Chang,Krovi,Chru,Haikka1} and approaches
\cite{Jing,Koch,Wu} of non-Markovian processes. Experimentally, the
simulation \cite{Xu1,Xu2} of non-Markovian environment has been
realized.

The measure of non-Markovianity of quantum evolution is a
fundamental problem which aims to detect whether a quantum process
is non-Markovian and how much degrees it deviates from a Markovian
one. Based on the distinguishability of quantum states, Breuer,
Laine and Piilo (BLP) \cite{Breuer} proposed a measure to detect the
non-Markovianity of quantum processes which is linked to the flow of
information between the system and environment. Alternatively,
Rivas, Huelga and Plenio \cite{Rivas} (RHP) also presented a measure
of non-Markovianity by employing the dynamical divisibility of a
trace-preserving completely positive map. It is clear that the BLP
measure is based on the physical features of the system-reservoir
interactions, while the RHP definition is based on the mathematical
property of the dynamical maps. It has been shown that the two
measures agree for several important and commonly-used models
\cite{Zeng}, but do not agree in general \cite{Dariusz}. In this
paper, we will use both the two measures to describe the
non-Markovianity of the dynamics of the considered system, so as to
more clearly see their relation, as well as the rationality as the
measure of non-Markovianity.

The study of the dynamics of non-Markovian open quantum systems is
typically very involved and often requires some approximations.
Almost all the previous treatments are based on the RWA, that is,
neglect the counter-rotating terms in the microscopic
system-reservoir interaction Hamiltonian. However, the
counter-rotating terms which are responsible for the virtual
exchanges of energy between the system and the environment not
always can be neglected. For example, for the wide-frequency-spectra
reservoir or when the frequency distribution of the structured
environment is detuned large enough from the transition of the
system, the RWA is invalid. Another motivation of this paper is thus
to study the effect of the counter-rotating terms on the
non-Markovian dynamics of the considered open quantum system.

The article is organized as follows. In Sec. II we introduce the
microscopic Hamiltonian model between the system and its
environment, and derive the non-Markovian time-local master equation
for a two-level system weakly coupled to a vacuum reservoir, by
using the TCL approach to the forth-order but without employing RWA
in the interaction Hamiltonian. In Sec. III, we investigate the
non-Markovianity of the system dynamics in terms of both the RHP and
BLP measures. Through the analytical solution in the secular
approximation, we obtain the sufficient and necessary conditions for
the dynamical indivisibility and the backflow of information,
showing the effect of the counter-rotating terms on the
non-Markovian dynamics of the system, and exposing the relations
between the BLP and RHP measures. In sec. IV, by choosing the
Lorentzian spectra reservoir as an exemplary example, we further
demonstrate the effect of the counter-rotating terms on the
dynamical indivisibility and the backflow of information, and
clarify the rationality of the two non-Markovian measures. Finally
in Sec.V, we discuss simply the complete positivity of the system
dynamics. And the conclusion is arranged in Sec.VI.

\section{The microscopic model}
Consider a two-level atom with Bohr frequency $\omega_{0}$
interacting with a zero-temperature bosonic reservoir modeled by an
infinite chain of quantum harmonic oscillators. The total
Hamiltonian for this system in the Schr\"{o}dinger picture is given
by
\begin{equation}
    H=\frac{1}{2}\omega_{0}\sigma_{z}+\sum_{k}\omega_{k}b^{+}_{k}b_{k}
    +\sum_{k}g_{k}(\sigma_{+}+\sigma_{-})(b_{k}+b^{+}_{k}),
\end{equation}
where $\sigma_{z}$ and $\sigma_{\pm}$ are the Pauli and inversion
operators of the atom, $\omega_{k}$, $b_{k}$ and $b_{k}^{+}$ are
respectively the frequency, annihilation and creation operators for
the $k$-th harmonic oscillator of the reservoir. The coupling
strength $g_{k}$ is assumed to be real for simplicity. The distinct
feature of this Hamiltonian is the reservation of the
counter-rotating terms, $\sigma_{+}b_{k}^{+}$ and $\sigma_{-}b_{k}$,
which is the so-called without RWA we call in this paper. Note however that our starting point is the dipole interaction
Hamiltonian between the atom and its environment, whose derivation
starting with the canonical Hamiltonian involves the discarding of a
term which is quadratic with respect to the radiation field. The discarding is not based on the RWA, but the fact that for low-intensity radiation, the quadratic term is much small
compared to the dipole interaction one \cite{Claude}.

The time-convolutionless projection operator technique is most
effective in dealing with the dynamics of open quantum systems. In
the limit of weak coupling between the system and the environment,
by expanding the TCL generator to the forth order with respect to
coupling strength, the non-Markovian master equation describing the
evolution of the reduced system, in the interaction picture, can be written as [For the main clue of its derivation, see appendix A.]
\begin{equation}
    \frac{d\rho(t)}{dt}=-i[H_{LS}(t),\rho(t)]+D[\rho(t)]+D'[\rho(t)],
\end{equation}
where
\begin{equation}
    H_{LS}(t)=S_{+}(t)\sigma_{+}\sigma_{-}+S_{-}(t)\sigma_{-}\sigma_{+},
\end{equation}
is the Lamb shift Hamiltonian which describes a small shift in the
energy of the eigenstates of the two-level atom. In many theoretical
researches \cite{Haikka}, this term was neglected usually. But in
this paper, we will take it into the consideration. The Lamb shift
includes the second and forth order contributions,
\begin{equation}
    S_{\pm}(t)=S_{\pm}^{II}(t)+S_{\pm}^{IV}(t),
\end{equation}
which respectively come from the second and forth order perturbative
expansion of the TCL generator. The second order Lamb shift is
\begin{equation}
    S_{\pm}^{II}(t)=\pm\int^{t}_{0}d\tau\int d\omega J(\omega)
    \sin[(\omega_{0}\mp\omega)\tau],
\end{equation}
with $J(\omega)=\sum_{k}|g_{k}|^{2}\delta(\omega-\omega_{k})$ the
spectral distribution of the environment. The expression for the
forth order Lamb shift $S_{\pm}^{IV}(t)$ is cumbersome which is
presented in the appendix A.

The dissipator $D[\rho(t)]$ that describes the
secular motion of the system has the form
\begin{eqnarray}
  D[\rho(t)] &=& \Gamma_{-}(t)\textbf{\{}\sigma_{-}\rho(t)\sigma_{+}-\frac{1}{2}\{\sigma_{+}\sigma_{-},\rho(t)\}\textbf{\}} \\
  \nonumber &+& \Gamma_{+}(t)\textbf{\{}\sigma_{+}\rho(t)\sigma_{-}-\frac{1}{2}\{\sigma_{-}\sigma_{+},\rho(t)\}\textbf{\}} \\
  \nonumber &+& \Gamma_{0}(t)\textbf{\{}\sigma_{+}\sigma_{-}\rho(t)\sigma_{+}\sigma_{-}-\frac{1}{2}\{\sigma_{+}\sigma_{-},\rho(t)\}\textbf{\}},
\end{eqnarray}
where the first line describes the dissipation of the atom to the
vacuum environment with time-dependent decay rate $\Gamma_{-}(t)$,
and the second line denotes the heating of the atom in the vacuum
environment with time-dependent heating rate $\Gamma_{+}(t)$. This
heating is related to the dissipation, for a ground-state atom in a
zero-temperature environment, there is no heating effect.
Dissipation and heating are usually accompanied by decoherence. The
last line in eq.(6) describes the pure decoherence with
time-dependent decoherence rate $\Gamma_{0}(t)$. The time-dependent
transition rates $\Gamma_{\pm}(t)$ also include the second and forth
order perturbative contributions of the TCL generator,

\begin{equation}
    \Gamma_{\pm}(t)=\Gamma_{\pm}^{II}(t)+\Gamma_{\pm}^{IV}(t),
\end{equation}
with the second order contribution as
\begin{equation}
    \Gamma_{\pm}^{II}(t)=2\int_{0}^{t}d\tau\int d\omega J(\omega)
    \cos[(\omega_{0}\pm\omega)\tau].
\end{equation}
While $\Gamma_{0}(t)$ completely comes from the forth-order
perturbative contribution. All the forth-order contributions are
presented in the appendix A. Eq.(6) indicates that the dissipative
model of eq.(1), except for inducing the energy exchange between the
system and its environment, also makes decoherence of the system.
But the rate of decoherence is much less than that of energy dissipation,
because $\Gamma_{0}(t)$ is only a forth-order contribution term of
TCL perturbative expansion.

The dissipator $D'[\rho(t)]$ represents the contribution of the
so-called nonsecular terms, that is, terms oscillating rapidly with
Bohr frequency $\omega_{0}$,
\begin{equation}
    D'[\rho(t)]=[\alpha(t)+i\beta(t)]\sigma_{+}\rho(t)\sigma_{+}+h.c.,
\end{equation}
here $h.c.$ denotes the Hermitian conjugation. These nonsecular
terms sometimes may also be neglected under the so-called secular
approximation \cite{Maniscalco}. The time-dependent coefficients $\alpha(t)$
and $\beta(t)$ also include the second and forth order
contributions,
\begin{equation}
    \alpha(t)=\alpha^{II}(t)+\alpha^{IV}(t),
\end{equation}
\begin{equation}
    \beta(t)=\beta^{II}(t)+\beta^{IV}(t),
\end{equation}
with
\begin{equation}
    \alpha^{II}(t)=2\int_{0}^{t}d\tau\int d\omega J(\omega)
    \cos[\omega(t-\tau)]\cos[\omega_{0}(t+\tau)],
\end{equation}
and
\begin{equation}
    \beta^{II}(t)=2\int_{0}^{t}d\tau\int d\omega J(\omega)
    \cos[\omega(t-\tau)]\sin[\omega_{0}(t+\tau)].
\end{equation}
The forth-order contributions are listed in the appendix A.

Note that the dynamics for a two-level system embedded in a
zero-temperature structured environment, under RWA, can be solved exactly, where the corresponding master
equation has the Lindblad-like form \cite{Breuer3},
\begin{equation}
\frac{d}{dt}\rho(t)=-\frac{i}{2}S(t)[\sigma_{+}\sigma_{-}, \rho(t)]+\gamma(t)\textbf{\{}\sigma_{-}\rho(t)\sigma_{+}-\frac{1}{2}\{\sigma_{+}\sigma_{-},\rho(t)\}\textbf{\}},
\end{equation}
where the time-dependent decay rate $\gamma(t)$ and Lamb shift $S(t)$ are related to the correlation function of the reservoir. Comparing this equation with eq.(2), we
see that the last two terms in the dissipator $D[\rho(t)]$, that
is, the heating and the pure decoherence terms, as well as the nonsecular dissipator $D'[\rho(t)]$ and
the Lamb shift $S_{-}(t)$, are completely from the contribution of the counter-rotating
terms presented in the interaction Hamiltonian. While the decay rate
$\Gamma_{-}(t)$ and the Lamb shift $S_{+}(t)$ include the
contributions of both rotating and counter-rotating terms, but the
main contributions [i.e., the second-order terms
$\Gamma^{II}_{-}(t)$ and $S^{II}_{+}(t)$] come from the rotating terms.
In fact, by expanding the decay rate $\gamma(t)$ and the Lamb shift $S(t)$ to the second order with respect to coupling strength, one obtain $\Gamma^{II}_{-}(t)$ and $S^{II}_{+}(t)$ \cite{Breuer3}. In the following, we will show that the contributions that come from
the counter-rotating terms are important, in particular to the
short-time-scale non-Markovian behaviors.

\section{Measures of Non-Markovianity}
Recently, people have been interested in the study of non-Markovianity of open quantum systems. Several definitions or measures \cite{Breuer,Rivas,Wolf,Usha,Lu} of non-Markovian dynamics have been presented.
In this section, we will employ two of the measures, i.e., the RHP \cite{Rivas} and BLP \cite{Breuer} measures, to investigate the non-Markovian dynamics
of the considered system so as to see the effect of the counter-rotating terms on non-Markovianity and the relation between the two measures.

\subsection{Divisible and indivisible dynamics}
A trace-preserving completely positive map $\varepsilon(t_{2},0)$ that
describes the evolution from times zero to $t_{2}$ is divisible if it satisfies composition law,
\begin{equation}
    \varepsilon(t_{2},0)=\varepsilon(t_{2},t_{1})\varepsilon(t_{1},0),
\end{equation}
with $\varepsilon(t_{2},t_{1})$ being completely positive for any $t_{2}\geq t_{1}\geq 0$. Due to the continuity of time, eq.(15) is always fulfilled in form. The key point for divisibility is actually the complete positivity of $\varepsilon(t_{2},t_{1})$ for any $t_{2}\geq t_{1}\geq 0$. If there exist times $t_{1}$ and $t_{2}$ such that the map $\varepsilon(t_{2},t_{1})$ is not completely positive, then the dynamical map $\varepsilon(t_{2},0)$ is indivisible. RHP \cite{Rivas} defined all the divisible maps to be Markovian. Therefore, the indivisibility of
a map advocates its dynamical non-Markovianity. It was shown that all the evolutions governed by Lindblad-type master equation with positive
transition rates are divisible \cite{Alicki}, thus Markovian.

It was proved \cite{Rivas} that the indivisibility of map $\varepsilon(t,0)$ is equivalent to the complete positivity of the quantity,
\begin{equation}
g(t)=\lim_{\epsilon\rightarrow0^{+}}\frac{\|[\varepsilon(t+\epsilon,t)\otimes I]|\Phi\rangle\langle\Phi|\|-1}{\epsilon}.
\end{equation}
Only for divisible map, $g(t)=0$. Where $|\Phi\rangle$ is a maximally entangled
state between the system of interest and an ancillary particle, and the map $\varepsilon$ performs
only on the state of the system.
Using the time-local master equation $\frac{d\rho}{dt}=\mathcal{L}_{t}(\rho)$, this expression
may be equivalently written as \cite{Rivas}
\begin{equation}
g(t)=\lim_{\epsilon\rightarrow0^{+}}\frac{\|[I+(\mathcal{L}_{t}\otimes
I)\epsilon]|\Phi\rangle\langle\Phi|\|-1}{\epsilon}.
\end{equation}
The function $g(t)$ is the so-called RHP non-Markovian measure. If and only if $g(t)=0$ for every time $t\in \{0,t_{2}\}$, the map $\varepsilon(t_{2},0)$ is Markovian. Otherwise it is non-Markovian. The distinctive advantage of RHP non-Markovian measure is that its calculation can be processed only by the use of time-local
master equation, not requiring the exact form of the dynamical map $\varepsilon(t,0)$.
In the following, we call the time interval that satisfies $g(t)>0$ the indivisible dynamical interval (IDI). For a non-Markovian process, there must exist one or several IDIs.

For the open two-level system considered in this paper, suppose that $|\Phi\rangle=\frac{1}{\sqrt{2}}[|01\rangle+|10\rangle]$, a straightforward deduction using equations (2) and (17) gives

\begin{eqnarray}
g &=& \frac{1}{4}|\Gamma_{-}+\Gamma_{+}+\sqrt{(\Gamma_{-}-\Gamma_{+})^{2}+4(\alpha^{2}+\beta^{2})}| \\
 \nonumber &+&\frac{1}{4}|\Gamma_{-}+\Gamma_{+}-\sqrt{(\Gamma_{-}-\Gamma_{+})^{2}+4(\alpha^{2}+\beta^{2})}|
 \\
 \nonumber
 &+&\frac{1}{4}[|\Gamma_{0}|-\Gamma_{0}-2\Gamma_{-}-2\Gamma_{+}],
\end{eqnarray}
where for compactness we omit the argument of all the time-dependent coefficients.
Obviously, the Lamb shift $H_{LS}(t)$ has no effect on the
indivisibility of the system dynamics.

\subsection{Backflow of information}
The second measure of non-Markovianity for quantum processes of open
systems we employ is proposed by BLP \cite{Breuer} which is based on
the consideration in purely physics. Note that Markovian processes
always tend to continuously reduce the trace distance between any
two states of a quantum system, thus an increase of the trace
distance during any time interval implies the emergence of
non-Markovianity. BLP further linked the change of the trace
distance to the flow of information between the system and its
environment, and concluded that the back flow of information from
environment to the system is the key feature of a non-Markovian
dynamics. In quantum information science, the trace distance for
quantum states $\rho_{1}$ and $\rho_{2}$ is defined as
\cite{Nielsen}
\begin{equation}
D(\rho_{1},\rho_{2})=\frac{1}{2}tr|\rho_{1}-\rho_{2}|,
\end{equation}
with $|A|=\sqrt{A^{+}A}$. For a given pair of initial states
$\rho_{1,2}(0)$ of the system, the change of the dynamical
trace-distance can be described by its time derivative
\begin{equation}
\sigma\textbf{(}t,\rho_{1,2}(0)\textbf{)}=\frac{d}{dt}D\textbf{(}\rho_{1}(t),\rho_{2}(t)\textbf{)},
\end{equation}
where $\rho_{1,2}(t)$ are the dynamical states of the system
with the initial states $\rho_{1,2}(0)$. For Markovian
processes, the monotonically reduction of the trace distance implies
$\sigma\textbf{(}t,\rho_{1,2}(0)\textbf{)}\leq 0$ for any initial states
$\rho_{1,2}(0)$ and at any time $t$. If there exists a pair of
initial states of the system such that for some evolutional time
$t$, $\sigma\textbf{(}t,\rho_{1,2}(0)\textbf{)}> 0$, then the information takes
backflow from environment to the system, and the process is non-Markovian.

In order to calculate the BLP measure, we must solve the dynamics of the
system. For this purpose, we write the alternative Bloch equation of
eq.(2) as [see appendix B for their derivation],

\begin{eqnarray}
  \dot{b}_{x} &=& -\frac{1}{2}(\Gamma_{-}+\Gamma_{+}+\Gamma_{0}-2\alpha)b_{x}+(S_{-}-S_{+}-\beta) b_{y}, \\
  \dot{b}_{y} &=& -\frac{1}{2}(\Gamma_{-}+\Gamma_{+}+\Gamma_{0}+2\alpha)b_{y}-(S_{-}-S_{+}+\beta) b_{x}, \\
  \dot{b}_{z} &=&
  -(\Gamma_{-}+\Gamma_{+})b_{z}+\Gamma_{+}-\Gamma_{-},
\end{eqnarray}
where the three components of the Bloch vector are defined as $
b_{j}(t)=\texttt{Tr}[\rho(t)\sigma_{j}]$ with $j=x,y,z$ and
$\sigma_{j}$ the Pauli operators. In terms of Bloch vector, the
trace distance of eq.(19) may be expressed as
\begin{equation}
    D(t)=\frac{1}{2}\sqrt{(\Delta b_{x})^{2}+(\Delta b_{y})^{2}+(\Delta b_{z})^{2}}
\end{equation}
where $\Delta b_{j}=b_{1j}(t)-b_{2j}(t)$ are the differences between the two Bloch components at evolutional time $t$. Correspondingly, the
derivative of this trace distance becomes
\begin{eqnarray}
\sigma &=& -\frac{1}{4}[(\Delta b_{x})^{2}+(\Delta b_{y})^{2}+(\Delta b_{z})^{2}]^{-1/2}\textbf{\{}(\Gamma_{-}+\Gamma_{+}+\Gamma_{0}-2\alpha)(\Delta b_{x})^{2} \\
 \nonumber &+&(\Gamma_{-}+\Gamma_{+}+\Gamma_{0}+2\alpha)(\Delta
 b_{y})^{2}+4\beta (\Delta b_{x})(\Delta
 b_{y})+2(\Gamma_{-}+\Gamma_{+})(\Delta b_{z})^{2}\textbf{\}},
\end{eqnarray}
where we have used the Bloch eqs.(21)-(23) in the deduction process. According BLP's criterion, $\sigma>0$ indicates the backflow of information from environment to the system. In the following, we call the time intervals in which $\sigma (t)>0$ the information-backflow intervals (IBIs).

\subsection{Secular approximation}
In order to see the effect of counter-rotating terms and make a distinct comparison between the BLP and RHP
measures in the current system, we now consider the case where the nonsecular term
$D'[\rho(t)]$ can be neglected, i.e., performing the so-called secular approximation. Here for the sake of discrimination, we call as in many literatures \emph{the rotating-wave approximation that used after tracing over the
bath degrees of freedom} the secular approximation. In other words, the secular approximation and the RWA have the same mathematical approaches--throwing away the rapidly oscillating terms in time, merely the times the approximations taking place are different. Just as pointed out in reference \cite{Maniscalco},
this kind of secular approximation though also is an average over rapidly oscillating terms, it does
not wash out the effect of the counter-rotating terms present in the
coupling Hamiltonian. Under the secular
approximation, the master equation (2) has the Lindblad-like
form with time-dependent transition rates, $\Gamma_{\pm}(t)$,
$\Gamma_{0}(t)$ and Lamb shift $H_{LS}(t)$. Employing the method
proposed in \cite{Michael}, the corresponding Bloch eqs.(21)-(23) in
this case can be solved exactly which gives
\begin{eqnarray}
  b_{x}(t) &=& e^{-\Theta(t)}[b_{x}(0)\cos\delta (t)-b_{y}(0)\sin\delta(t)], \\
  b_{y}(t) &=& e^{-\Theta(t)}[b_{x}(0)\sin\delta (t)+b_{y}(0)\cos\delta(t)],\\
  b_{z}(t) &=& e^{-\Lambda(t)}\left\{b_{z}(0)+\int_{0}^{t}ds e^{\Lambda(s)}[\Gamma_{+}(s)-\Gamma_{-}(s)]\right\},
\end{eqnarray}
with
\begin{equation}
\Theta(t)=\frac{1}{2}\int_{0}^{t}ds[\Gamma_{-}(s)+\Gamma_{+}(s)+\Gamma_{0}(s)],
\end{equation}
\begin{equation}
\Lambda(t)=\int_{0}^{t}ds[\Gamma_{-}(s)+\Gamma_{+}(s)],
\end{equation}
and
\begin{equation}
\delta(t)=\int_{0}^{t}ds[S_{+}(s)-S_{-}(s)].
\end{equation}
Inserting these solutions into eq.(25), we get
\begin{eqnarray}
\sigma &=& -\frac{1}{4}I(t)\textbf{\{}e^{-2\Theta(t)}(\Gamma_{-}+\Gamma_{+}+\Gamma_{0})[\textbf{(}\Delta b_{x}(0)\textbf{)}^{2}+\textbf{(}\Delta b_{y}(0)\textbf{)}^{2}]
 \\
 \nonumber
 &+&2e^{-2\Lambda(t)}(\Gamma_{-}+\Gamma_{+})\textbf{(}\Delta b_{z}(0)\textbf{)}^{2}\textbf{\}},
\end{eqnarray}
where $I(t)=\{e^{-2\Theta(t)}[\textbf{(}\Delta
b_{x}(0)\textbf{)}^{2}+\textbf{(}\Delta
b_{y}(0)\textbf{)}^{2}]+e^{-2\Lambda(t)}\textbf{(}\Delta
b_{z}(0)\textbf{)}^{2}\}^{-1/2}$ is a positive function and $\Delta
b_{j}(0)=b_{1j}(0)-b_{2j}(0)$ is the difference between the two
initial Bloch components. This expression shows that the sufficient
and necessary conditions for the backflow of information from
environment to the system are
\begin{equation}
  \Gamma_{-}(t)+\Gamma_{+}(t)+\Gamma_{0}(t)<0,
\end{equation}
or
\begin{equation}
    \Gamma_{-}(t)+\Gamma_{+}(t)<0.
\end{equation}
Because if at some time $t$, one of this conditions is satisfied,
then we can always find a pair of initial states such that $\sigma
(t)>0$. For example, if eq.(33) fulfils, it suffices to choose the
initial states satisfying $\Delta b_{z}(0)=0$. Conversely, if
$\sigma (t)>0$ at some time $t$, then at least one of the two conditions must
be satisfied.

On the other hand, under secular approximation, eq.(18) is simplified as
\begin{equation}
    g=\frac{1}{4}\{2|\Gamma_{-}|+2|\Gamma_{+}|+|\Gamma_{0}|-2\Gamma_{-}-2\Gamma_{+}-\Gamma_{0}\}.
\end{equation}
Obviously, when one of the three rate functions, $\Gamma_{-}(t)$, $\Gamma_{+}(t)$ or $\Gamma_{0}(t)$, is negative, then $g>0$, vice versa. Thus the sufficient and necessary conditions for the indivisibility of the dynamics are
\begin{equation}
    \Gamma_{-}(t)<0, or\hspace{0.2cm} \Gamma_{+}(t)<0, or\hspace{0.2cm} \Gamma_{0}(t)<0.
\end{equation}

Eqs. (33), (34) and (36) demonstrate two important results. One the one hand, the counter-rotating terms [which induce $\Gamma_{+}(t)$, $\Gamma_{0}(t)$ and a part of $\Gamma_{-}^{IV}(t)$] may have important effect to the non-Markovian dynamics of the system, according to RHP and BLP measures. On the other hand, they show that the conditions for the backflow of information are much more rigorous than that of indivisibility. The later only requires one of the transition rates to be negative, while the former further requires the sum of the two or the total transition rates to be negative. This conditionality once again validates the previous results: The backflow of information must lead the indivisibility of the dynamics, but the  reverse is not true \cite{Dariusz}. However, for Lindblad-like master equation with only single transition rate, the sufficient and necessary conditions for the two measures become clearly the same, denoting the consistency of the two measures in this case \cite{Zeng}.

\section{Non-Markovian dynamics for Lorentzian spectrum}
In order to further demonstrate quantitatively the effect of the counter-rotating terms, as well as the rationality of the two non-Markovian measures, we
specify our study to a particular reservoir spectra, Lorentzian
spectra,
\begin{equation}
J(\omega)=\frac{\gamma_{0}\lambda^{2}}{2\pi[(\omega_{0}-\omega-\Delta)^{2}+\lambda^{2}]},
\end{equation}
which describes the interaction of an atom with an imperfect cavity
and is widely used in literatures. Where $\omega_{0}$ denotes the
transition frequency of the atom, $\Delta=\omega_{0}-\omega_{c}$ is
the frequency detuning between the atom and the cavity mode.
$\lambda$ is the width of Lorentzian distribution, which is connected
to the reservoir correlation time $\tau_{R}=\lambda^{-1}$. The
parameter $\gamma_{0}$ can be regarded as the decay rate for the
excited atom in the Markovian limit of flat spectrum which is
related to the relaxation time $\tau_{S}=\gamma_{0}^{-1}$. For the
Lorentzian spectra, all the time-dependent coefficients including
$S_{\pm}(t)$, $\Gamma_{\pm}(t)$, $\Gamma_{0}(t)$, $\alpha(t)$ and
$\beta(t)$ can be calculated analytically, but the expressions are
too complicated. We thus study them only numerically.

In Fig.1, we show the time evolution of these coefficients. For our
purpose, we intentionally choose three special sets of parameters.
It shows that for narrow spectrum and small detuning [In Fig.1 (a)
and (d), $\lambda/\omega_{0}=0.2\% $, $\Delta/\omega_{0}=2\% $],
$\Gamma_{-}(t)$ plays the dominant role, while $\Gamma_{+}(t)$ and
$\Gamma_{0}(t)$ are almost zero. The nonsecular coefficients
$\alpha(t)$ and $\beta(t)$ in this case behave fast oscillations
[Fig.1 (d)], so that on average in time the effect can also be neglected.
These results imply that for this set of parameters, the
counter-rotating terms in Hamiltonian (1) play little effect
actually to the system dynamics and the commonly-used RWA is valid. However, for wider spectrum or/and larger
detuning, the results are different [see Fig.1 (b) and (c)], where
though $\Gamma_{0}$ is still near zero \cite{Note}, $\Gamma_{+}$
clearly can not be neglected. Thus the counter-rotating terms in
these cases are important and the RWA is invalid. Of course, for very wide spectrum, one
may expect that the dynamics tends to be Markovian. The positivity
of the $\Gamma_{\pm}(t)$ and $\Gamma_{0}(t)$ in Fig.1 (c) confirms
this point. In addition, when $\lambda$ is small, the correlation time
of the environment is longer, thus $\Gamma_{-}$ in Fig.1 (a)
oscillates to emerge negative values in a relatively longer time.
With the increasing of $\lambda$, the correlation time becomes small
and small, and the times for $\Gamma_{\pm}$ to be negative shorten
or even vanish [Fig.1 (b) and (c)]. Note that the observable
negative values of $\Gamma_{+}$ in Fig.1 (b) demonstrate the
contribution of the counter-rotating terms to the non-Markovianity
of the system dynamics.

\begin{figure}[b]
\includegraphics{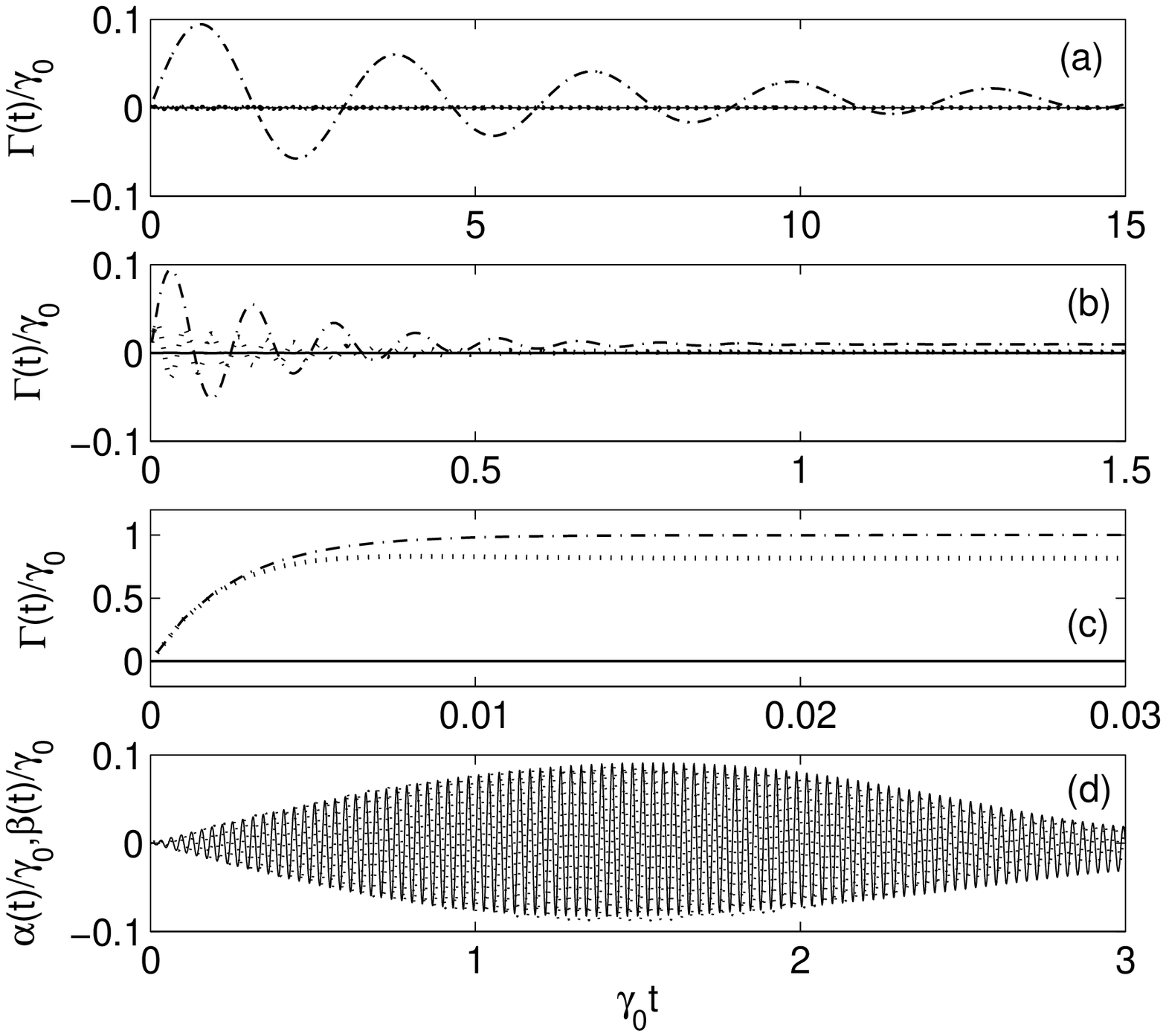}
\caption{\label{fig:epsart} Evolution of the
time-dependent coefficients. The dot-dash line, dot line and solid
line in (a),(b) and (c) correspond to respectively the evolutions
of $\Gamma_{-}$, $\Gamma_{+}$ and $\Gamma_{0}$, while the solid and dot
lines in (d) refer to the evolutions of $\alpha$ and $\beta$. Where
we choose $\omega_{0}=100\gamma_{0}$, and $\lambda=0.2\gamma_{0}$,
$\Delta=2\gamma_{0}$ for (a); $\lambda=5\gamma_{0}$,
$\Delta=50\gamma_{0}$ for (b); $\lambda=400\gamma_{0}$,
$\Delta=10\gamma_{0}$ for (c). The parameters for (d) are the same
as that of (a).}
\end{figure}

Note that in the RWA, the corresponding master equation (14) may be solved exactly. For the Lorentzian
spectrum, the RHP and BLP measures may be expressed as \cite{Zeng},
\begin{equation}
g(t)=\left\{%
\begin{array}{lll}
    0 & \mathrm{for} &\gamma(t)\geq0 \\
    -\gamma(t) &\mathrm{for} & \gamma(t)<0\\
\end{array}%
\right.
\end{equation}
and
\begin{equation}
\sigma(t)=-\gamma(t)F(t).
\end{equation}
where
\begin{equation}
\gamma(t)=\texttt{Re}\left[\frac{2\gamma_{0}\lambda\sinh(dt/2)}{d\cosh(dt/2)+(\lambda-i\Delta)\sinh(dt/2)}\right],
\end{equation}
with $d=\sqrt{(\lambda-i\Delta)^{2}-2\gamma_{0}\lambda}$. The positive real function $F(t)$ is defined as,
\begin{equation}
F(t)=\frac{a^{2}e^{-\frac{3}{2}\Gamma(t)}+|b|^{2}e^{-\frac{1}{2}\Gamma(t)}}
{\sqrt{a^{2}e^{-\Gamma(t)}+|b|^{2}}},
\end{equation}
with $\Gamma(t)=\int_{0}^{t}dt'\gamma(t')$, and
$a=\langle1|\rho_{1}(0)|1\rangle-\langle1|\rho_{2}(0)|1\rangle$,
$b=\langle1|\rho_{1}(0)|0\rangle-\langle1|\rho_{2}(0)|0\rangle$ being
the differences of the population and of coherence respectively
 for the two given initial states. Eqs.(38) and (39) show that under the RWA, the distributions of IDIs and IBIs are exactly the same, which are determined by $\gamma(t)<0$. In the following, we study numerically the evolution of the measures $\sigma$ and $g$, under the condition without using RWA, so as to further highlight the non-Markovian effect of the counter-rotating terms, as well as the rationality of BLP and RHP measures.

In Fig.2, we show the time evolution of the measure $\sigma$ in the same parameters as in Fig.1, where the solid lines are plotted according to eq.(25) and the dot lines according to eq.(39). We choose the pair of initial states to be $\rho_{1}(0)=|1\rangle\langle1|$ and $\rho_{2}(0)=|0\rangle\langle0|$, which can maximize the BLP measure \cite{Breuer}. For evidence, we only give the time intervals of $\sigma>0$, i.e., the IBIs. We can see clearly the corrections of the counter-rotating terms on the BLP measure. In Fig.2 (a), both the distributions of the IBIs and the shapes of the two curves are similar, responding that the counter-rotating terms make lesser effect to the backflow of information in this case which is in line with the idea of RWA. The dips on each peaks of the solid-line in Fig.2 (b) are due to the negativity of $\Gamma_{+}(t)$ at that times [see Fig.1 (b)], implying that $\Gamma_{+}$ has the offset on the backflow of information. There is no IBI in Fig.2 (c), denoting that under the choice of this set of parameters, there is no backflow of information, or equivalently the dynamics is Markovian according to BLP measure, which is in line with the non-negativity of $\Gamma_{\pm}(t)$ and $\Gamma_{0}(t)$. In addition, the time scale for the backflow of information is consistent with the reservoir correlation time $\lambda^{-1}$ [Fig.2 (a), (b)]. All these results show that on the one hand the counter-rotating terms can affect the backflow of information, and on the other hand the correction of the counter-rotating terms on the backflow of information is reasonable.

\begin{figure}[b]
\includegraphics{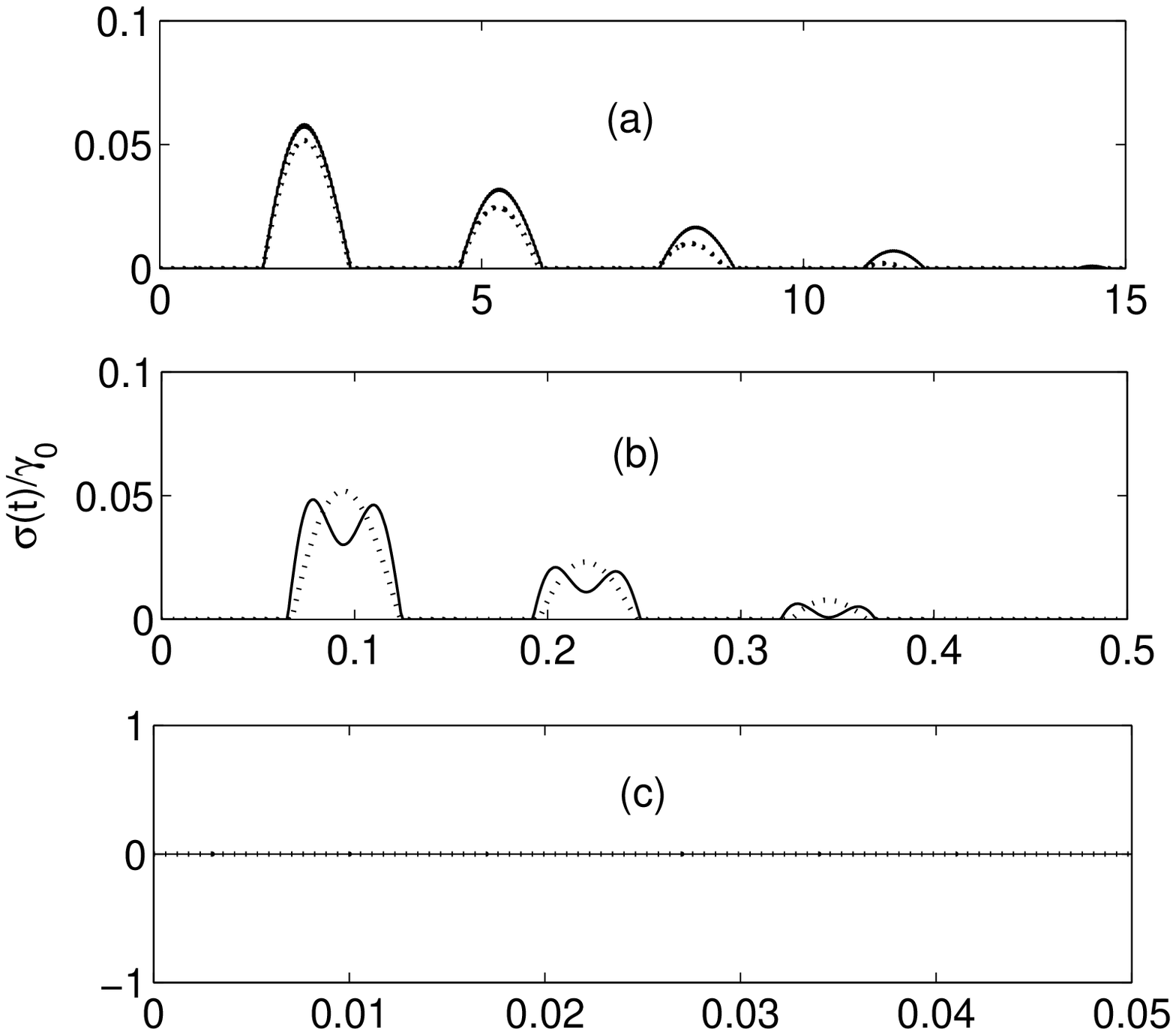}
\caption{\label{fig:epsart} Time evolution of $\sigma(t)$,
with the solid and dot lines corresponding to respectively eqs.(25)
and (39). The parameters in (a),(b) and (c) are set to be in accordance with that in Fig.1.}
\end{figure}


In Fig.3, we plot the time evolution of the measure $g$ in the same parameters as in Fig.1. We see that when the counter-rotating terms are omitted, the distribution of the IDIs agrees with that of IBIs [see the dot lines in Figs. 2 and 3]. The non-Markovian time scale predicted by measure $g$ is also in accordance with the reservoir correlation time $\lambda^{-1}$. The horizontal dot line in Fig.3(c) denotes that under the choice of those parameters, the dynamics is actually Markovian. All these results show that with no counter-rotating terms, the RHP and BLP measures agree. Both of them can depict rightly the non-Markovianity of the underlying dynamics. However, when the counter-rotating terms are considered, the case is distinctly different: The IDIs now become $(0,\infty)$ [see the solid lines in Fig.3], which are clearly inconsistent with the practice. Because first of all, the non-Markovian time scales in the underlying conditions are never infinite. Next, in Fig.3(a), the choice of the parameters is consistent with the RWA, the result after considering counter-rotating terms should has some tiny, not distinct amendments, over the result under RWA. For the parameters in Fig.3(c), the reservoir correlation time $\lambda^{-1}$ is very short and the system dynamics is actually Markovian, should not appearing long-time non-Markovianity. These egregious results denote that the RHP measure in these cases is invalid. Note that the reason for resulting in these unpractical phenomena is mainly due to the nonsecular coefficients $\alpha(t)$ and $\beta(t)$. When these nonsecular coefficients are neglected, eq.(35) is not seen to deviate obviously from the practice.

\begin{figure}[b]
\includegraphics{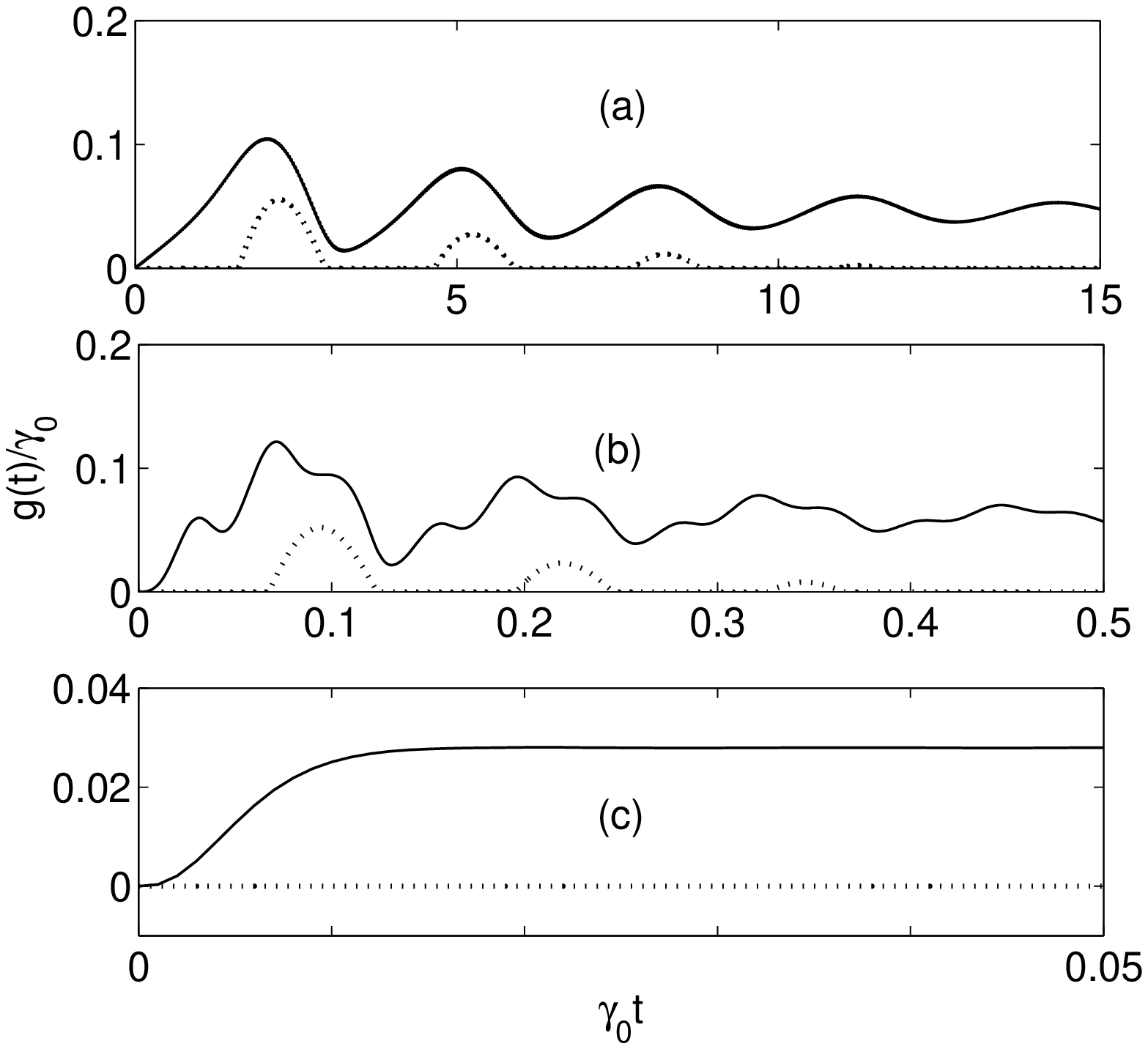}
\caption{\label{fig:epsart} Time evolution of $g(t)$,
where the solid and dot lines are plotted according to eqs.(18) and
(38) respectively. The parameters are set to be the same as in Fig.1.}
\end{figure}


\section{Complete positivity}
The evolution of a real physical state should be not only positive
but also complete positive. In practical theoretical study, however,
due to the application of some assumptions and approximations, the
positivity or the complete positivity may not always be satisfied.
Here we present a study of the complete positivity for our
considered model, i.e., the master equation of (2). As the damping
matrix has the block diagonal form (see appendix B), thus we can
directly use the conditions for complete positivity presented by
Hall \cite{Michael}. The necessary condition of the complete
positivity, for the master equation (2), may be given by two
inequalities:
\begin{equation}
    \Lambda(t)\geq0,
\end{equation}
\begin{equation}
    2\Theta(t)\geq\Lambda(t),
\end{equation}
with $\Theta(t)$ and $\Lambda(t)$ given by eqs.(29)-(30). The
sufficient condition is also given by two inequalities. The first
one coincides with eq.(42) and the second one may be expressed as
\begin{equation}
    \chi(t)
    \cosh\theta(t)\leq1+A^{2}(t)-\kappa^{2}(t)-2|A(t)-\chi(t)|,
\end{equation}
where $\chi(t)=e^{-2\Theta(t)}$, $A(t)=e^{-\Lambda(t)}$,
$\kappa(t)=A(t)\int_{0}^{t}ds[\Gamma_{+}(s)-\Gamma_{-}(s)]A^{-1}(s)$,
 and $\theta(t)=2\int_{0}^{t}ds\sqrt{\alpha^{2}(s)+\beta^{2}(s)}$
 with $\alpha(t)$, $\beta(t)$ given by eqs.(10)-(11). Using
 inequality (43) to release the modulus in the right-hand side, we
 get
\begin{equation}
    \chi(t)
    \cosh\theta(t)\leq[1-A(t)]^{2}+2\chi(t)-\kappa^{2}(t),
\end{equation}
 Note
 that the left-hand side of eq.(45) is relevant to the nonsecular motion,
 but the right-hand side only depends on the secular motion. As $\theta(t)$ increases with time $t$,
 eq.(45) is not satisfied for long times. But in short non-Markovian time scales we are interested in, it may be fulfilled.
In order to see this, we plot the time evolution of function
$G(t)\equiv[1-A(t)]^{2}+2\chi(t)-\kappa^{2}(t)-\chi(t)\cosh\theta(t)$
as in Fig.4, for the same parameters as in Fig.1 and under the
Lorentzian spectra. Obviously, in the scale of the correlation times
$\lambda^{-1}$, $G(t)>0$. The condition of eqs.(42)-(43) is
satisfied for all times in this case. Thus in the short
non-Markovian time scales, the evolution of the system is physical.
In the secular regime, the sufficient condition eq.(45) can be
relaxed to
 \begin{equation}
    [1-A(t)]^{2}+\chi(t)-\kappa^{2}(t)\geq 0,
 \end{equation}
which can be satisfied for much more longer times for the Lorentzian reservoir.

\begin{figure}[b]
\includegraphics{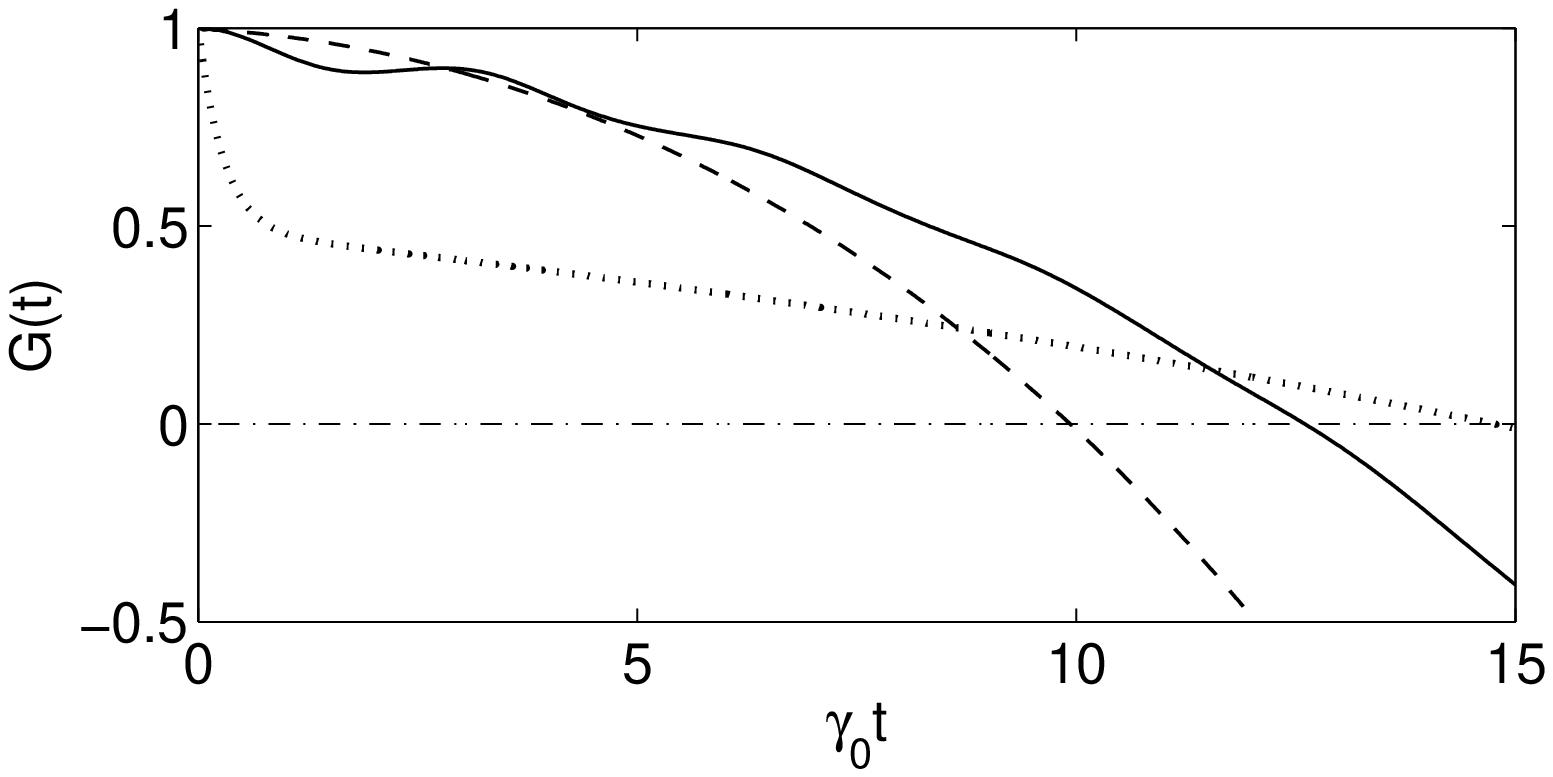}
\caption{\label{fig:epsart} Time evolution of $G(t)$, where the solid, dash and dot lines
correspond to respectively the parameters in Fig.1 (a), (b) and
(c).}
\end{figure}

\section{Conclusion}

In conclusion, we have studied the non-Markovianity of the dynamics for a two-level
system interacting with a zero-temperature structured environment
without using RWA. In the limit of weak coupling
between the system and its reservoir, by expanding the TCL generator to the forth order with respect to the coupling
strength, we have derived the time-local
non-Markovian master equation for the reduced state of the system.
Under the secular approximation, the TCL master equation has the Lindblad-like form with time-dependent transition rates. We have obtained the exact analytic solution.
The sufficient and necessary conditions for the indivisibility and the backflow
of information for the system dynamics were presented, which showed two
important results: First, the counter-rotating terms may play important
roles to the indivisibility and the backflow of information for the system
dynamics. Second, it showed explicitly that the BLP and RHP measures generally do not coincide. It demonstrated more clearly the previous result:
The backflow of information must lead to the indivisibility of dynamics, but the reserve is not true.

When the nonsecular terms are included, we have investigated numerically
the non-Markovian properties of the system dynamics by assuming that the
environment spectrum is Lorentzian.
By compared with the result under RWA, we found
that the BLP measure is corrected appropriately, but the RHP measure is inconsistent with practice, showing
that the RHP measure has finite applicable range.

Finally, we have discussed the complete positivity of the underlying dynamics. We have presented the sufficient and necessary
conditions of the complete positivity. Numerical simulation showed
that these conditions can be satisfied in the short non-Markovian
time scale.

The measure of non-Markovianity is a fundamental problem in the
study of open quantum system dynamics. Although several measures of non-Markovianity have been presented already, it is noted that these measures are not completely equivalent to
each other. Therefore, the problem for measuring the non-Markovianity
of quantum processes still remains elusive and, in some sense,
controversial. At present stage, it is meaningful and necessary to expose the characteristics of various measures and their relations in some concrete systems.

The investigation of a two-level system interacting with a bath of harmonic oscillators,
i.e., the spin-boson model, is of particular interest in the theory of open quantum system.
In the context of quantum computation, it represents a qubit coupled to an environment,
which can produce dissipation and decoherence.
Though in the numerical simulations we have only considered the Lorentzian environment,
our analytic results adapt to other structured environments, such as the Ohmic reservoir,
the photonic band-gap material \cite{Woldeyohannes}, etc. By properly engineering the structure of the environment,
one can control the non-Markovian dynamics of the open quantum system,
so as to effectively control the evolution of some interesting physical quantities, such as the quantum coherence,
quantum entanglement and discord, etc. Therefore, our work will be helpful for the quantum information processing.

Of course, our model is not fully general. First of all,
we have considered only a two-level system weakly coupled a zero-temperature environment.
Next, our starting point is based on the dipole interaction Hamiltonian between the atom and its environment, not on the canonical Hamiltonian.
Finally, we have used the TCL perturbation expansion for the derivation of master equation eq.(2). Thus our results are still conditional and further investigations may be necessary.

\begin{acknowledgments}
This work is supported by the National
Natural Science Foundation of China (Grant No.11075050), the
National Fundamental Research Program of China (Grant
No.2007CB925204), the Program for Changjiang Scholars and Innovative Research Team in University under Grant No.IRT0964, and the Construct Program of the
National Key Discipline.
\end{acknowledgments}

\appendix

\section{Derivation of the master equation and the time-dependent forth-order coefficients}

In our study, the derivation of the forth-order TCL master equation
(2) is very cumbersome. Here we can present only the main clue about
the deduction. Our calculation is based on the description of
reference \cite{Breuer3} about the TCL projection operator
technique. By assuming a factoring initial condition
$\rho(0)=\rho_{S}(0)\otimes\rho_{B}$ for the system and environment,
one obtains a homogeneous TCL master equation [see (9.33) of
\cite{Breuer3}]
\begin{equation}
    \frac{\partial}{\partial
    t}\mathcal{P}\rho(t)=\mathcal{K}(t)\mathcal{P}\rho(t).
\end{equation}
Due to the assumption of vacuum reference state
$\rho_{B}=|0\rangle\langle0|$ for the environment, the TCL generator
$\mathcal{K}(t)$ only has even-order terms in its perturbation
expansion. The second- and forth-order TCL generators may be
calculated directly via eqs.(9.61)-(9.62) of reference
\cite{Breuer3}, where the related operators $F_{k}$ and $Q_{k}$ in
the interaction picture are given by,
\begin{eqnarray}
  F_{k}(t) &=& \sigma_{+}e^{i\omega_{0}t}+\sigma_{-}e^{-i\omega_{0}t}, \\
  Q_{k}(t) &=&
  g_{k}(b_{k}e^{-i\omega_{k}t}+b_{k}^{+}e^{i\omega_{k}t}).
\end{eqnarray}
Calculating the second- and forth-order TCL generators and sorting
them in operators, then eq.(A1) reduces to the required master equation.

In the master equation (2), each of the time-dependent coefficients
consists of in principle two parts--the second and the forth order
parts. The second-order parts have relatively simple expressions,
but the expressions of the forth-order parts are very complex. In
terms of abbreviation $t_{ij}=t_{i}-t_{j}$ with $t_{0}\equiv t$,
$C(t)=\int d\omega J(\omega)\cos\omega t$, $ S(t) = \int d\omega
J(\omega)\sin\omega t$ and
$\texttt{T}\int=\int_{0}^{t}dt_{1}\int_{0}^{t_{1}}dt_{2}\int_{0}^{t_{2}}dt_{3}$,
the forth-order coefficients may be written in the following,
\begin{eqnarray}
  S_{+}^{IV}(t) &=& 2 \texttt{T}\int\textbf{\{}[S(t_{02})\sin(\omega_{0}t_{03})- 3C(t_{02})\cos(\omega_{0}t_{03})]C(t_{13})\sin(\omega_{0}t_{12}) \\
  \nonumber &+& [C(t_{02})\sin(\omega_{0}t_{03})- S(t_{02})\cos(\omega_{0}t_{03})]S(t_{13})\sin(\omega_{0}t_{12}) \\
  \nonumber &+& [S(t_{03})\sin(\omega_{0}t_{02})- 3C(t_{03})\cos(\omega_{0}t_{02})]C(t_{12})\sin(\omega_{0}t_{13}) \\
  \nonumber &+& [C(t_{03})\sin(\omega_{0}t_{02})- S(t_{03})\cos(\omega_{0}t_{02})]S(t_{12})\sin(\omega_{0}t_{13}) \\
  \nonumber &+& [-S(t_{03})\sin(\omega_{0}t_{01})- C(t_{03})\cos(\omega_{0}t_{01})]C(t_{12})\sin(\omega_{0}t_{23}) \\
  \nonumber &+& [-C(t_{03})\sin(\omega_{0}t_{01})+ S(t_{03})\cos(\omega_{0}t_{01})]S(t_{12})\sin(\omega_{0}t_{23})\textbf{\}},
\end{eqnarray}

\begin{eqnarray}
  S_{-}^{IV}(t) &=& 2\texttt{T}\int\textbf{\{}[S(t_{02})\sin(\omega_{0}t_{03})+ C(t_{02})\cos(\omega_{0}t_{03})]C(t_{13})\sin(\omega_{0}t_{12}) \\
  \nonumber &+& [C(t_{02})\sin(\omega_{0}t_{03})- S(t_{02})\cos(\omega_{0}t_{03})]S(t_{13})\sin(\omega_{0}t_{12}) \\
  \nonumber &+& [S(t_{03})\sin(\omega_{0}t_{02})+ C(t_{03})\cos(\omega_{0}t_{02})]C(t_{12})\sin(\omega_{0}t_{13}) \\
  \nonumber &+& [C(t_{03})\sin(\omega_{0}t_{02})- S(t_{03})\cos(\omega_{0}t_{02})]S(t_{12})\sin(\omega_{0}t_{13}) \\
  \nonumber &+& [-S(t_{03})\sin(\omega_{0}t_{01})- C(t_{03})\cos(\omega_{0}t_{01})]C(t_{12})\sin(\omega_{0}t_{23}) \\
  \nonumber &+& [-C(t_{03})\sin(\omega_{0}t_{01})+ S(t_{03})\cos(\omega_{0}t_{01})]S(t_{12})\sin(\omega_{0}t_{23})\textbf{\}},
\end{eqnarray}

\begin{eqnarray}
  \Gamma_{\pm}^{IV}(t) &=& -8\texttt{T}\int\textbf{\{}[C(t_{13})\sin(\omega_{0}t_{03})\pm S(t_{13})\cos(\omega_{0}t_{03})]C(t_{02})\sin(\omega_{0}t_{12}) \\
  \nonumber &+& [C(t_{12})\sin(\omega_{0}t_{02})\pm S(t_{12})\cos(\omega_{0}t_{02})]C(t_{03})\sin(\omega_{0}t_{13}) \\
  \nonumber &\mp& [S(t_{03})C(t_{12})+ C(t_{03})S(t_{12})]\sin(\omega_{0}t_{23})\cos(\omega_{0}t_{01})\textbf{\}},
\end{eqnarray}

\begin{eqnarray}
  \Gamma_{0}(t) &=& 16\texttt{T}\int\textbf{\{}[C(t_{02})C(t_{13})+S(t_{02})S(t_{13})]\sin(\omega_{0}t_{03})\sin(\omega_{0}t_{12}) \\
  \nonumber &+& [C(t_{03})C(t_{12})+S(t_{03})S(t_{12})]\sin(\omega_{0}t_{02})\sin(\omega_{0}t_{13})  \\
  \nonumber &+& [C(t_{03})C(t_{12})-S(t_{03})S(t_{12})]\sin(\omega_{0}t_{01})\sin(\omega_{0}t_{23})\textbf{\}},
\end{eqnarray}

\begin{eqnarray}
  \alpha^{IV}(t) &=& -8\texttt{T} \int\textbf{\{}S(t+t_{2})S(t_{13})\sin\omega_{0}(t+t_{3})\sin(\omega_{0}t_{12}) \\
  \nonumber &+& S(t+t_{3})S(t_{12})\sin\omega_{0}(t+t_{2})\sin(\omega_{0}t_{13}) \\
  \nonumber &+& [C(t_{03})C(t_{12})- S(t_{03})S(t_{12})]\sin\omega_{0}(t+t_{1})\sin(\omega_{0}t_{23})\textbf{\}},
\end{eqnarray}

\begin{eqnarray}
  \beta^{IV}(t) &=& 8\texttt{T}\int\textbf{\{}S(t_{02})S(t_{13})\cos\omega_{0}(t+t_{3})\sin(\omega_{0}t_{12}) \\
  \nonumber &+& S(t_{03})S(t_{12})\cos\omega_{0}(t+t_{2})\sin(\omega_{0}t_{13}) \\
  \nonumber &+& [C(t_{03})C(t_{12})- S(t_{03})S(t_{12})]\cos\omega_{0}(t+t_{1})\sin(\omega_{0}t_{23})\textbf{\}}.
\end{eqnarray}

\section{Derivation of Bloch equation}

According to the definition of Bloch vector $
b_{j}(t)=\texttt{Tr}[\rho(t)\sigma_{j}]$, we have $
\dot{b}_{j}(t)=\texttt{Tr}[\dot{\rho}(t)\sigma_{j}]$. By inserting
master equation (2) into it and after some deduction, one can obtain
the required Bloch equation. For example, for the component equation
concerning $\dot{b}_{x}$ we have,
\begin{equation}
\dot{b}_{x}(t)=-i\texttt{Tr}\{[H_{LS}(t),\rho(t)]\sigma_{x}\}+\texttt{Tr}\{D[\rho(t)]\sigma_{x}\}+
\texttt{Tr}\{D'[\rho(t)]\sigma_{x}\}.
\end{equation}
By use of the circulation property of trace operation and the
Pauli algorithm, one easily get
$-i\texttt{Tr}\{[H_{LS}(t),\rho(t)]\sigma_{x}\}=(S_{-}-S_{+})b_{y}$,
$\texttt{Tr}\{D[\rho(t)]\sigma_{x}\}=-\frac{1}{2}(\Gamma_{-}+\Gamma_{+}+\Gamma_{0})b_{x}$,
and $\texttt{Tr}\{D'[\rho(t)]\sigma_{x}\}=\alpha b_{x}-\beta b_{y}$.
Summing up them, we thus obtain eq.(21).

The Bloch eqs.(21)-(23) can also be written as the compact vector
form, $\dot{\textbf{b}}=M\textbf{b}+\textbf{v}$, with the damping
matrix $M$ and drift matrix $\textbf{v}$ given respectively by
\begin{equation}
      M=\left(
              \begin{array}{ccc}
                -\frac{1}{2}(\Gamma_{-}+\Gamma_{+}+\Gamma_{0}-2\alpha) & S_{-}-S_{+}-\beta & 0 \\
                -(S_{-}-S_{+}+\beta) & -\frac{1}{2}(\Gamma_{-}+\Gamma_{+}+\Gamma_{0}+2\alpha) & 0 \\
                0 & 0 & -(\Gamma_{-}+\Gamma_{+}) \\
              \end{array}
            \right),
\end{equation}
and $\textbf{v}^{T}=\left(
                      \begin{array}{ccc}
                        0, & 0, & \Gamma_{+}-\Gamma_{-} \\
                      \end{array}
                    \right).
$ Note that the damping matrix is in block diagonal form.


\begin{thebibliography}{99}
\bibitem{Note} The reason for $\Gamma_{0}$ always very much smaller than $\Gamma_{\pm}$ is that the former is the forth-order small quantity of the TCL generator, while the later are the second-order small quantities.
\end{thebibliography}
\end{document}